Investigation of 3D-Quantized Ring Vortices in Rotating Coupled Atom-Molecular BEC

Sreoshi Dutta[1], Krishna Rai Dastidar[2] and Chanchal Chaudhuri[3*]

[1] Dept. of Physics, Vivekananda College, Thakurpukur, Kolkata-700063, *India.*

[2] School of physical Sciences, Indian Association for the Cultivation of Science, Kolkata-700032, India.

[3] Dept. of Physics, University of Gour Banga, Malda-732103, West Bengal, India.

*Abstract:*

We study the coupled atom-molecular quantized ring vortices of $^{87}$Rb Bose-Einstein Condensates (BEC) trapped in a rotating three dimensional (3D) anisotropic cylindrical trap both in time independent and time-dependent Gross-Pitaevskii approaches. For atom to molecular conversion and vice-versa two-photon Raman photoassociation scheme has been used. Atomic and molecular stationary state solutions show that the different number of nodes and crests formed in the density profile (as a function of r and z) for different combinations of radial (**n**) and axial (**n$_z$**) quantum numbers at a fixed azimuthal quantum number **l**=2, give rise to different structure around the 3D ring vortex centered at r=0. We have considered both spontaneous and induced decays and compared the results with those without considering the decays. The out of phase oscillation of atomic and molecular numbers in two vortex states both in presence and absence of external decays is the signature of coherence due to the atom-molecular coupling. This coherence is also implemented in the evolution of coupled atomic and molecular vortices. Intensity of molecular ring vortices grows with progress in time in expense of that of atomic ring vortices and vice versa. It is found that the intensity of the coupled atomic and molecular ring vortices starts oscillation out of phase during evolution. Dependence of the atom-molecular conversion efficiency and the lifetime of the system on the laser intensity of photoassociation lasers and the total number of atoms in two different vortex states reveals that formation of atom-molecular coupled vortices and the efficiency of formation can be controlled by varying these parameters. Linear stability analysis of vortex states as a function of different system parameters shows that the atomic vortices are more stable than molecular vortices and the stable atomic and molecular vortices can be achieved by controlling these parameters.

I. **Introduction:**

Experimental observations of the atomic Bose-Einstein Condensate (BEC) in trapped dilute gases [1] using laser light has led to many experimental and theoretical investigations on atomic BEC systems [1-5]. Experimental observations of the atomic vortex and its characterizations have been performed by different groups [6]. A number of theoretical work have beendone to investigate atomic vortex [7] such as spontaneous shape deformation leading to the formation of a vortex, dynamics of vortex formation in merging BE-condensate fragments, 'hidden' vortices in a rotating double well potential, the nucleation of spontaneous vortices in trapped Fermi gases undergoing a BCS-BEC crossover, 3D atomic vortex solitons, dynamics of single and multiple ring vortices, vortex-antivortex pairing in decaying superfluids, vortex formation in dipolar BEC, and recently active investigations on turbulence in trapped BEC. Dalfovo's group has shown that after releasing the trap the core of quantized atomic vortices expands faster than the atomic cloud of BEC which will facilitate the observation of atomic vortices, by solving the Gross-Pitaevski equation numerically and compared it with analytical results. Consequently investigations on two component BECs were started. Mathews *et al.* observed vortices in two-component Bose-Einstein condensates and explored differences in the dynamics and stability of vortices [8]. Extensive theoretical studies [9] have been done to explore different aspects of two-component BE condensates e.g. considering dipole-dipole interaction, both attractive and repulsive atom-atom interactions, spin-orbit (SO)-coupling etc using the Gross-Pitaevskii equation.

With the advent of studies on photoassociation processes in cold atoms and atomic BECs [10,11], investigations on the study of atom-molecular coupled BEC systems have started. Formation of molecules from atomic BEC or ultracold atoms by photoassociation was first introduced by Julienne's group [10] through a two-step process involving stimulated free-bound transition followed by spontaneous bound-bound emission where the later results in the production of incoherent mixture of a large range of vibrational levels in the electronic ground state of the molecules. Hence stimulated bound-bound transition was chosen to achieve the state-specific population of the final molecular state of interest either through an adiabatic or a nonadiabatic pathway, which is essentially a stimulated two-photon Raman photoassociation. Experimentally state selective molecules at rest has been created from atomic BEC by photoassociation process adopting the Raman two-photon stimulated free-bound and bound-bound transition technique [12,13] and by magnetoassociation process applying Feshbach resonance technique [14-16]. This technique has also been used for condensation of molecular Fermi gases [17]. Many theoretical [18-24] attempts were also made for understanding and setting the guidelines for realisation of molecular BECs using two-photon Raman photoassociation as well as magnetic Feshbach resonances. A number of theoretical efforts have been made to find out the enhancement of conversion efficiency through different approaches such as time-dependent magnetic field in conjunction with stimulated Raman transition [23], making of proper tuning of the pulse duration [24, 25] and the coherent population trapping in Feshbach resonance-assisted stimulated Raman adiabatic pathway [26, 27]. Rai Dastidar's group [21] explored the coherences in the evolution of atomic and molecular density due to the atom-molecular coupling via Raman two-photon association and by magnetic Feshbach resonance using a modified Gross-Pitaevskii approach in coupled atom-molecular BEC system and the nature of out of phase oscillation of atomic and molecular density has been compared with experimental results.

The structure of vortices in two species or two-component atomic condensates and its dependence on the system parameters have been studied [8, 9]. However, the studies on the formation of vortices in atom-molecular coupled BEC system are different from that of two component atomic BEC systems due to the presence of atom-molecular conversion coupling and the mass of molecules is twice that of atoms in the former. Some theoretical investigations have been done on the formation, structure and stability of vortices and vortex lattices in atom-molecular coupled BEC system. Julienne's group analyzed structure and stability of vortices in hybrid atomic-molecular BEC using Gross-Pitaevskii model adopting the stimulated Raman-induced photoassociation process [28]. They predicted new types of topological vortex states in coherently coupled two-component condensates even without a trap, and demonstrated the nontrivial dynamics of the coupled system in the presence of losses. Bigelow's group has studied effect of atom-molecular coupling on the formation and structure of vortex lattices in rotating atom-molecular coupled BEC system [29]. The structural phase transitions in this coupled system have been explored by studying the dependence of degree of phase matching on the system parameters e.g. atom-molecular coupling strength, atom-molecular interaction and rotation frequency of coupled system. Liu's group has explored dependence of nature of vortices on different combinations of quantum numbers, principal and secondary quantum numbers in rotating atom-molecular coupled BEC system by using analytical solutions for vortices [30]. They have studied formation of vortex lattices in atom-molecular coupled BEC system considering space modulated nonlinearity and dependence of structure of vortex lattices on the Raman detuning and atom-molecular coupling strength. They have shown that atom-molecular interaction plays a crucial role to control the structure of vortex lattices. However to our knowledge no experimental study on coupled atom-molecular vortices has been done yet.

In the present study we have investigated the dynamics, stability and control over the formation of atomic and molecular coupled vortices in a rotating coupled atom-molecular BEC of $^{87}$Rb atoms. In this scheme coherent $^{87}$Rb$_2$ molecules are obtained in the lowest vibrational level of the electronic ground state via two-photon Raman photoassociation (PA) of $^{87}$Rb atoms. Here Gross-Pitaevskii equation (GPE) with additional rotational term has been used. For the formation of coupled atom molecular quantized vortices in the steady state time independent 3D coupled GP equations have been solved using imaginary time method. Dependence of the structure and shape of the atomic and molecular vortices on the different combinations of radial and axial quantum numbers (due to the presence of nodes and crests in the density profiles) have been studied. To study the dynamics of these coupled vortices in 3D we have solved time-dependent coupled GP equations using Crank-Nicholson method. It has been shown previously that atomic and molecular density in (non-rotating) coupled atom molecular BEC system in the ground state oscillate coherently due to the presence of atom-molecular conversion coupling [19, 21]. In this work we have explored the signature of coherences in the dynamics of atomic and molecular quantized three dimensional ring vortices (which are the excited states of the coupled system) in rotating atom-molecular coupled BEC system. The feasibility of formation and the stability of coupled atom-molecular quantized vortices have been investigated by choosing different system parameters which are experimentally realizable.

In practice external decays (spontaneous and induced) may play a crucial role on the stability of the coupled system. Hence spontaneous decay of excited atoms and the decay of molecules in two different channels (bound and the continuum of the ground state) induced by PA lasers have been considered. The dependence of conversion efficiency and the decay time of the total number of particles to 1/e times of its initial value on the initial number of atoms and the laser intensities have been studied considering the external decays. To investigate the stability of atomic and molecular vortices formed in different vortex states the imaginary part of the energy of the vortex states of atoms and molecules has been studied as a function of system parameters e.g. initial number of atoms, laser intensity, atom-molecular interaction and the rotation frequency of the coupled system.

In this paper theoretical framework has been described in section II: Theoretical formulation is given in subsection IIA and the numerical methods have been described in subsection IIB. Results and the discussions have been given in section III. Results on the formation of vortices in the steady state have been given in subsection IIIA and the results on the dynamics of atom molecular quantized vortices have been described in subsection IIIB. The variation of decay time of the system, atom to molecule conversion efficiency and the dependence of stability of vortex states on different system parameters have been discussed in subsection IIIC. Finally the conclusions have been drawn in section IV.

**II. Theoretical Framework**:

**IIA. Theoretical Model:**

Fig.1 schematically shows the two-photon Raman photoassociation scheme where two atoms initially in the state |i> of total energy $2E_1$ collide to form a molecule in the rovibrational state |v> of energy $E_v$ of an excited electronic state in the presence of the coupling laser field of frequency $\omega_1$, which subsequently undergoes a transition to the ground state |g> of energy $E_2$ through stimulated emission by the second coupling laser of frequency $\omega_2$. The two-photon Raman detuning is given as $\delta = (2E_1 - E_2)/\hbar - (\omega_2 - \omega_1)$ and the two-photon transition will be resonant when $\delta$ equals to zero.

In order to study the evolution of atomic and molecular vortices in rotating coupled atomic and molecular BECs, we solved the coupled Gross-Pitaevskii equations (GPE) of motion for atoms (wavefunction $\psi_a$) and molecules (wavefunction $\psi_m$) as follows.

$$i\hbar \frac{\partial \psi_a}{\partial t} = \left[ -\frac{\hbar^2 \nabla^2}{2m} + U_a(r,z) + \lambda_a |\psi_a|^2 + \lambda_{am} |\psi_m|^2 \right] \psi_a$$
$$+ \chi \psi_a \psi_m^* - i\hbar\alpha\psi_a - \hbar\beta_1\psi_a - i\hbar\Gamma_1|\psi_a|^2\psi_a$$
$$+ i\hbar\Omega \frac{\partial \psi_a}{\partial \theta} \tag{1}$$

$$i\hbar \frac{\partial \psi_m}{\partial t} = \left[ -\frac{\hbar^2 \nabla^2}{4m} + U_m(r,z) + \lambda_m |\psi_m|^2 + \lambda_{am} |\psi_a|^2 \right.$$
$$\left. - \hbar\delta \right] \psi_m + \frac{\chi}{2} \psi_a \psi_a^* - \hbar\beta_2\psi_m - i\hbar\Gamma_2\psi_m$$
$$+ i\hbar\Omega \frac{\partial \psi_m}{\partial \theta} \tag{2}$$

where $U_a(r,z) = \frac{1}{2}m\omega_r^2(r^2 + \lambda^2 z^2)$ and $U_m(r,z) = m\omega_r^2(r^2 + \lambda^2 z^2)$ are the external anisotropic cylindrical harmonic trap potentials for atoms and molecules respectively, $\lambda$ is the anisotropic factor $(= \frac{\omega_z}{\omega_r})$, where $\omega_r$ and $\omega_z$ are the angular trapping frequencies along radial and axial direction of the system respectively. $\lambda_a$, $\lambda_m$, $\lambda_{am}$ are the atom-atom, molecule-molecule and atom-molecule interaction strength respectively. According to Bogoliubov mean field theory $\lambda_a = \frac{4\pi\hbar^2 a}{m}$, where $a$ is the s-wave scattering length for atom-atom interaction and we consider that $\lambda_a = \lambda_m = \lambda_{am}$ for simplicity. The spontaneous decay rate is denoted by α and $\Gamma_1$, $\Gamma_2$ indicate induced decay rates, $\beta_1$, $\beta_2$ represent the atomic and molecular light shift terms. Furthermore, $\tilde{\delta}$ is related to effective two-photon Raman detuning ($\tilde{\delta} = \delta + \beta_2 - 2\beta_1$) and Ω is the angular frequency of rotation of the system. The last terms in eqns. (1) and (2) represent the rotational energy terms producing vortices. χ is the atom-molecular Raman coupling constant, i.e., the conversion factor from atom to molecule and vice-versa.

The form of atom-molecular Raman coupling is expressed as

$$\frac{\chi}{\hbar} = -\frac{\Omega_1 \Omega_2}{2\sqrt{2}} \sum_v \frac{I_{1,v} I_{2,v}^*}{\Delta_v} \qquad (3)$$

Where $\Omega_1$ is the free-bound and $\Omega_2$ is the bound-bound Rabi frequencies which are functions of laser intensities $I_1$ and $I_2$ respectively: $\Omega_{1,2} \propto \sqrt{I_{1,2}}$. The expression for the atom-atom scattering length can be written as

$$a = a_{bg} - \frac{m}{4\pi\hbar} \sum_v \left[ \frac{\Omega_1^2}{4\Delta_v} + \frac{\Omega_2^2}{4\Delta_v^{(1)}} \right] |I_{1,v}|^2 \tag{4}$$

Here $a_{bg}$ is the background scattering length, $I_{1,v}$ is the free-bound and $I_{2,v}$'s are the bound-bound Frank-Condon factors. The spontaneous decay rate from an excited state of an atom is

$$\alpha = \frac{\gamma_a}{8} \sum_{i=1,2} \frac{(D_i^A)^2}{D_i^2} \tag{5}$$

Two induced decay rates, atomic loss rates due to one photon association and the spontaneous Raman scattering rates for molecules are denoted as (stimulated)

$$\Gamma_j = \frac{\gamma_M}{8} \sum_v \left[ \frac{(\Omega_j)^2}{\Delta_v^2} + \frac{(\Omega_{3-j})^2}{\left(\Delta_v^{(j)}\right)^2} \right] |I_{j,v}|^2, (j = 1,2) \tag{6}$$

The atomic and molecular light-shift terms are represented as respectively

$$\beta_1 = \sum_{i=1,2} \frac{(D_i^A)^2}{4D_i} \tag{7}$$

$$\beta_2 = \sum_v \left[ \frac{(\Omega_2)^2}{4\Delta_v} + \frac{(\Omega_1)^2}{4\Delta_v^{(2)}} \right] |I_{2,v}|^2 \tag{8}$$

Here $\gamma_a$ and $\gamma_M$ are the spontaneous decay rates of atoms and molecules respectively, $D_i = \omega_0 - \omega_i$ are the detuning of laser from the resonant frequency $\omega_0$ of the atomic transition between the dissociation limit of the ground and excited state energy. The expression for the respective detuning factors are given below,

$$\Delta_v = (E_v - 2E_1)/\hbar - \omega_1 \tag{9}$$

$$\Delta_v^{(1)} = (E_v - 2E_1)/\hbar - \omega_2 \tag{10}$$

$$\Delta_v^{(2)} = \frac{E_v - 2E_1}{\hbar} - \omega_1 \tag{11}$$

The above coupled GP equations (1) and (2) reduce to dimensionless GP equations by rescaling the length by linear harmonic oscillator length $a_{HO} = \sqrt{\hbar/m\omega}$, energy by $\hbar\omega$ and time by $1/\omega$ in cylindrical polar coordinates system as follow:

$$i\frac{\partial \psi_a}{\partial t} = \left[-\frac{1}{2}(\frac{\partial^2}{\partial r^2} + \frac{1}{r}\frac{\partial}{\partial r} - \frac{l^2}{r^2} + \frac{\partial^2}{\partial z^2}) + \frac{1}{2}(r^2 + \lambda^2 z^2)\right.$$
$$\left. + g_a|\psi_a|^2 + g_{am}|\psi_m|^2\right]\psi_a + \frac{\chi}{\hbar}\psi_a\psi_m^* - i\alpha\psi_a$$
$$- \beta_1\psi_a - i\Gamma_1|\psi_a|^2\psi_a + i\Omega\frac{\partial \psi_a}{\partial \theta}$$

$$\tag{12}$$

$$i\frac{\partial \psi_m}{\partial t} = \left[-\frac{1}{4}\left(\frac{\partial^2}{\partial r^2} + \frac{1}{r}\frac{\partial}{\partial r} - \frac{l^2}{r^2} + \frac{\partial^2}{\partial z^2}\right) + (r^2 + \lambda^2 z^2)\right.$$

$$\left. + g_m|\psi_m|^2 + g_{am}|\psi_a|^2 + \epsilon\right]\psi_m + \frac{\chi}{2\hbar}\psi_a\psi_a^*$$

$$- \beta_2\psi_m - i\Gamma_2\psi_m + i\Omega\frac{\partial \psi_m}{\partial \theta} \qquad (13)$$

We assume the form of the wave function $\psi(r,z,\theta,t)$ both for atom and molecule as

$$\psi_i(r,z,\theta,t) = f_i(r,z)e^{il\theta - i\mu_i t} \qquad (14)$$

where i corresponds to atoms and molecules. $\mu_i$'s are the chemical potentials, '$l$' is azimuthal quantum number also known as *intrinsic vorticity* and $f_i(r,z)$'s are the time-independent wave-functions. Using the wave-function from equation (14) in equations (12) and (13) the time-independent forms of the GP equations take the following forms

$$\left[-\frac{1}{2}\left(\frac{\partial^2}{\partial r^2} + \frac{1}{r}\frac{\partial}{\partial r} - \frac{l^2}{r^2} + \frac{\partial^2}{\partial z^2}\right) + \frac{1}{2}(r^2 + \lambda^2 z^2) + g_a|f_a|^2\right.$$

$$\left. + g_{am}|f_m|^2\right]f_a + \frac{\chi}{\hbar}f_a f_m^* - l\Omega f_a = \mu_a f_a \qquad (15)$$

$$\left[-\frac{1}{4}\left(\frac{\partial^2}{\partial r^2}+\frac{1}{r}\frac{\partial}{\partial r}-\frac{l^2}{r^2}+\frac{\partial^2}{\partial z^2}\right)+(r^2+\lambda^2 z^2)+g_m|f_m|^2\right.$$
$$\left.+g_{am}|f_a|^2\right]f_m+\frac{\chi}{2\hbar}f_a f_a^* - l\Omega f_m = \mu_m f_m \qquad (16)$$

The approximate stationary state solutions of the equations (15) and (16) neglecting the nonlinearity of the system can be written as [31]

$$f_a(r,z) \sim r^l L_n^l(r^2) e^{-\frac{1}{2}(r^2+\lambda z^2)} H_{n_z}(\sqrt{\lambda}z) \qquad (17)$$

$$f_m(r,z) \sim r^l L_n^l(2r^2) e^{-(r^2+\lambda z^2)} H_{n_z}(\sqrt{2\lambda}z) \qquad (18)$$

where the functions $f_a$ and $f_m$ involve Gaussian-Laugurre-Hermite functions and the corresponding chemical potentials are (as $\lambda_a = \lambda_m$)

$$\mu_{a,m} = 2n + 1 + l - l\Omega + (n_z + \frac{1}{2})\lambda \qquad (19)$$

## IIB. Numerical Approach:

We obtained the numerical stationary state solutions for the atomic and molecular vortices in coupled BEC system by solving the time-independent atomic-molecular coupled GP equations (15) and (16) using imaginary time evolution method taking $f_a$ and $f_m$ as the initial wave functions. We reduce the nonlinear Schrödinger equations (NLSE) which have been used for our coupled system in its dimensionless form replacing radial distance $r = a_{HO}r'$, axial distance $z = a_{HO}z'$ and time $t = \tau/\omega$, where $a_{HO} = \sqrt{\hbar/m\omega}$ as follows:

$$\left[-\frac{1}{2}\left(\frac{\partial^2}{\partial r'^2} + \frac{1}{r'}\frac{\partial}{\partial r'} - \frac{l^2}{r'^2} + \frac{\partial^2}{\partial z'^2}\right) + \frac{1}{2}\left(r'^2 + \lambda^2 z'^2\right) + g_a|f_a|^2 \right.$$
$$\left. + g_{am}|f_m|^2\right]f_a + \chi' f_a f_m^* - l\Omega' f_a = \mu_a f_a \qquad (20)$$

$$\left[-\frac{1}{4}\left(\frac{\partial^2}{\partial r'^2} + \frac{1}{r'}\frac{\partial}{\partial r'} - \frac{l^2}{r'^2} + \frac{\partial^2}{\partial z'^2}\right) + \left(r'^2 + \lambda^2 z'^2\right) + g_m|f_m|^2 \right.$$
$$\left. + g_{am}|f_a|^2 + \epsilon'\right]f_m + \frac{\chi'}{2} f_a f_a^* - l\Omega' f_m = \mu_m f_m \qquad (21)$$

where $\mu_a$ and $\mu_m$ are the eigenvalues of the atomic and molecular time-independent NLSEs. In order to obtain the numerical solutions of the stationary state equations (20) and (21), we implement the imaginary time method as follows:

$$f_j(r',z',\tau + \Delta\tau) = (1 - i\Delta\tau H'_j)f_j(r',z',\tau) \qquad (22)$$

where $H'_j$ denotes the Hamiltonian corresponding to atomic and molecular coupled equation in its' dimensionless form. Numerical iteration is performed using the length step 0.01 along both the direction r and z within the range (0 to 5) and (-5 to 5) respectively. To get the converged solution for atomic and molecular vortices, the convergence of the solution has been checked to be $10^{-6}$ for the wave-function. The time step we have used in this imaginary time iteration is $5\times10^{-8}$ and at each time steps of imaginary time, total number of atom and double of number of molecules have been normalized to total number particles N i. e. $\iint(|f_a|^2 + 2|f_m|^2)2\pi r' dr' dz' = N$.

To study the dynamical behaviour of atomic and molecular vortices in coupled atomic-molecular system we have solved the time-dependent coupled GP equations by applying steepest descent method in Crank-Nicholson scheme. We assume the wave function is of the form $\psi = \phi/\sqrt{r}$ for the sake of simplicity of our cylindrical system and to solve the equations (12) and (13) we impose the boundary conditions which are $\phi_{a,m}(r,z,t) \to 0$ as $r, z \to 0$; and $|\phi_{a,m}(r,z,t)| \to e^{-(r^2+\lambda z^2)/2}$ as $r, z \to \infty$. Using Crank-Nicholson scheme we discretise coupled atomic and molecular time-dependent nonlinear Schrödinger equations (12) and (13) (in dimensionless form) in both the radial and axial directions and hence we obtain a pair of dimensionless equations for atomic BEC as follows:

$$i \frac{\phi_{a,j,k}^{n+\frac{1}{2}} - \phi_{a,j,k}^n}{\Delta \tau}$$

$$= -\frac{1}{4h^2} \left[ \left( \phi_{a,j+1,k}^{n+\frac{1}{2}} - 2\phi_{a,j,k}^{n+\frac{1}{2}} + \phi_{a,j-1,k}^{n+\frac{1}{2}} \right) \right.$$

$$+ \left( \phi_{a,j+1,k}^n - 2\phi_{a,j,k}^n + \phi_{a,j-1,k}^n \right) \bigg]$$

$$+ \left[ \frac{1}{2} r_j^2 + \frac{1}{2r_j^2}\left(l^2 - \frac{1}{4}\right) + g_a \frac{|\phi_{a,j,k}^n|^2}{r_j} \right.$$

$$+ g_{am} \frac{|\phi_{m,j,k}^n|^2}{r_j} - i\frac{\alpha}{2\omega} - \frac{\beta_1}{2\omega} - i\frac{\Gamma_1}{2\omega}\frac{|\phi_{a,j,k}^n|^2}{r_j}$$

$$\left. - l\Omega \right] \left( \frac{\phi_{a,j,k}^{n+\frac{1}{2}} + \phi_{a,j,k}^n}{2} \right) + \frac{\chi}{\omega} \frac{\phi_{a,j,k}^{n*} \phi_{m,j,k}^n}{\sqrt{r_j}} \qquad (23)$$

$$i \frac{\phi_{a,j,k}^{n+1} - \phi_{a,j,k}^{n+\frac{1}{2}}}{\Delta \tau}$$

$$= -\frac{1}{4h^2} \left[ \left( \phi_{a,j,k+1}^{n+1} - 2\phi_{a,j,k}^{n+1} + \phi_{a,j,k-1}^{n+1} \right) \right.$$

$$+ \left( \phi_{a,j,k+1}^{n+\frac{1}{2}} - 2\phi_{a,j,k}^{n+\frac{1}{2}} + \phi_{a,j,k-1}^{n+\frac{1}{2}} \right) \bigg]$$

$$+ \left[ \frac{1}{2}\lambda^2 z_k^2 + -i\frac{\alpha}{2\omega} - \frac{\beta_1}{2\omega} - l\Omega \right] \left( \frac{\phi_{a,j,k}^{n+1} + \phi_{a,j,k}^{n+\frac{1}{2}}}{2} \right) \qquad (24)$$

and a pair of dimensionless equations for molecular BEC as follows:

$$i\frac{\phi_{m,j,k}^{n+\frac{1}{2}} - \phi_{m,j,k}^{n}}{\Delta\tau}$$

$$= -\frac{1}{8h^2}\left[\left(\phi_{m,j+1,k}^{n+\frac{1}{2}} - 2\phi_{m,j,k}^{n+\frac{1}{2}} + \phi_{m,j-1,k}^{n+\frac{1}{2}}\right)\right.$$

$$\left. + \left(\phi_{m,j+1,k}^{n} - 2\phi_{m,j,k}^{n} + \phi_{m,j-1,k}^{n}\right)\right]$$

$$+ \left[r_j^2 + \frac{1}{4r_j^2}\left(l^2 - \frac{1}{4}\right) + g_m\frac{|\phi_{m,j,k}^{n}|^2}{r_j}\right.$$

$$+ g_{am}\frac{|\phi_{a,j,k}^{n}|^2}{r_j} + \frac{\delta}{\omega} - \frac{\beta_2}{2\omega} - i\frac{\Gamma_2}{2\omega}$$

$$\left. - l\Omega\right]\left(\frac{\phi_{m,j,k}^{n+\frac{1}{2}} + \phi_{m,j,k}^{n}}{2}\right) + \frac{\chi}{2\omega}\frac{\phi_{a,j,k}^{n*}\phi_{a,j,k}^{n}}{\sqrt{r_j}} \quad (25)$$

$$i\frac{\phi_{m,j,k}^{n+1} - \phi_{m,j,k}^{n+\frac{1}{2}}}{\Delta\tau}$$

$$= -\frac{1}{8h^2}\left[\left(\phi_{m,j,k+1}^{n+1} - 2\phi_{m,j,k}^{n+1} + \phi_{m,j,k-1}^{n+1}\right)\right.$$

$$\left. + \left(\phi_{m,j,k+1}^{n+\frac{1}{2}} - 2\phi_{m,j,k}^{n+\frac{1}{2}} + \phi_{m,j,k-1}^{n+\frac{1}{2}}\right)\right]$$

$$+ \left[\frac{1}{2}\lambda^2 z_k^2 + -i\frac{\alpha}{2\omega} - \frac{\beta_1}{2\omega} - l\Omega\right]\left(\frac{\phi_{m,j,k}^{n+1} + \phi_{m,j,k}^{n+\frac{1}{2}}}{2}\right) \quad (26)$$

where the discretized wavefunctions are $\phi_{i,j,k}^n = \phi_i(r_j, z_k, \tau_n); i = a, m,$ for atoms and molecules respectively. Here the dimensionless r', z' and τ' has been substituted by r, z and τ for simplicity.

For Numerical calculation the stationary state solution obtained numerically solving the time independent equations has been used as wavefunction in time dependent equation at t=0. The maximum length in radial and axial directions have been taken in units of $a_{HO}$ is 5. We use the step size (h) along radial and axial directions are the same 0.01 and the time step (Δτ) is 0.1μs. The dynamics of the coupled system has been studied for 0.3ms.

### III. Linear Stability Analysis:

For the linear stability analysis we consider the perturbed solution of equation (12) and (13) as follows

$$\Psi_a = \left[\Phi_a(r,z) + u_1 e^{i\lambda t} + w_1 e^{-i\lambda t}\right] e^{-i\mu_a t} \qquad (27)$$

$$\Psi_m = \left[\Phi_m(r,z) + u_2 e^{i\lambda t} + w_2 e^{-i\lambda t}\right] e^{-i\mu_m t} \qquad (28)$$

where $|u_1| \ll 1$, $|u_2| \ll 1$, $|w_1| \ll 1$ and $|w_2| \ll 1$ are the small perturbation neglecting the higher order term. Substituting these wave functions (27) and (28) in the equation (12) and (13) we get the following eigenvalue equation having eigenvalue λ.

$$\begin{bmatrix} L_1 & \sqrt{2}\chi\Phi_m - g_a\Phi_a^2 + i\Gamma_1\Phi_a^2 & \sqrt{2}\chi\Phi_a - g_{am}\Phi_a\Phi_m & -g_{am}\Phi_a\Phi_m \\ g_a\Phi_a^2 - i\Gamma_1\Phi_a^2 - \sqrt{2}\chi\Phi_m & -L_1 & g_{am}\Phi_a\Phi_m & g_{am}\Phi_a\Phi_m - \sqrt{2}\chi\Phi_m \\ \sqrt{2}\chi\Phi_a - g_{am}\Phi_a\Phi_m & -g_{am}\Phi_a\Phi_m & L_2 & -g_m\Phi_m^2 \\ g_{am}\Phi_a\Phi_m & g_{am}\Phi_a\Phi_m - \sqrt{2}\chi\Phi_a & g_m\Phi_m^2 & -L_2 \end{bmatrix} \begin{bmatrix} u_1 \\ w_1 \\ u_2 \\ w_2 \end{bmatrix}$$

$$= \lambda \begin{bmatrix} u_1 \\ w_1 \\ u_2 \\ w_2 \end{bmatrix} \qquad (29)$$

The matrix elements $L_1$ and $L_2$ are defined as

$$L_1 = \frac{1}{2}\left(\frac{\partial^2}{\partial r^2} + \frac{\partial^2}{\partial z^2} - \frac{\partial}{r\partial r}\right) - 2g_a\Phi_a^2 - g_{am}\Phi_m^2$$
$$-\frac{1}{2}(r^2 + \lambda^2 z^2) + \mu_a + l\Omega + i\alpha + \beta_1 \qquad (30)$$

$$L_2 = \frac{1}{4}\left(\frac{\partial^2}{\partial r^2} + \frac{\partial^2}{\partial z^2} - \frac{\partial}{r\partial r}\right) - 2g_m\Phi_m^2 - g_{am}\Phi_a^2$$
$$-\frac{1}{4}(r^2 + \lambda^2 z^2) + \mu_m + l\Omega - \delta + i\Gamma_2 + \beta_2 \qquad (31)$$

We solved the eigenvalue equation (29) to get the eigenvalues. If the eigenvalues are purely real then the wave functions are stable otherwise they are unstable.

### III. Results and Discussion:

In the two-photon Raman photoassociation scheme used here the formation of the $^{87}Rb_2$ molecules is considered to be in the lowest vibrational level of the ground state $^3\Sigma_u^+(V_g(R))$ via $0_g^-$ excited state $(V_e(R))$ by association of two $^{87}Rb$ atoms from the continuum of the ground state (Fig.1).

The parameters used in this calculations $\gamma_A = 3.7 \times 10^7 s^{-1}$, $\gamma_M = 2\gamma_A$, $\Omega_1 = 2.3 \times 10^{10} s^{-1}$ $\Omega_2 = 6.324 \times 10^9 s^{-1}$, $\chi/\hbar = 7.6 \times 10^{-7} m^{3/2} s^{-1}$, $\Gamma_1 = 1.629 \times 10^{-23} m^3 s^{-1}$, $\Gamma_2 = 304.4\ s^{-1}$, $\beta_1 = 2.108 \times 10^7 s^{-1}$, $\beta_2 = 3.344 \times 10^6 s^{-1}$, $\alpha = 134.06 s^{-1}$, $\delta = 3.879 \times 10^7 s^{-1}$ [19] unless otherwise mentioned. In this calculation the effective detuning delta-tilde (which includes the Ac Stark shift, $\beta_1, \beta_2$ and the two-photon Raman detuning, delta) has been kept fixed. The s-wave scattering length is 5.4 nm [21]. We have started with initial number of atoms N=50000 unless otherwise variation of N is considered. The frequency of cylindrically symmetric harmonic trap potential in radial direction has been taken as $\omega_r/2\pi = 100$ Hz whereas in the axial direction the frequency $\omega_z = 0.36\omega_r$ [21]. The angular frequency of rotation of the condensate system has been taken as $\Omega = 0.6\omega_r$. For the sake of simplicity we have considered the strengths of all three interactions, atom-atom, molecule-molecule and atom-molecule are equal, i.e., $\lambda_a = \lambda_m = \lambda_{am}$ unless otherwise dependence of results on interactions is considered.

### III A. Formation of Vortices:

To demonstrate the formation of quantized vortices in the rotating coupled atom-molecular condensate, we obtained the stationary state solutions of time-independent rotating coupled GP equations (15) and (16) numerically, using imaginary time evolution method taking the Gaussian-Laguerre-Hermite functions as initial wave-functions (eqns. (17) and (18)). The iteration has been continued until the convergence of wave functions reaches to $10^{-6}$.

Results of stationary state solutions have been shown in Fig.2 for different values of radial (n) and axial ($n_z$) quantum numbers for a fixed value of azimuthal quantum number (l). Fig.2(a) and Fig.2(b) show the plot of density profiles ($|\Phi_{a,m}(r,z)|^2$) as a function of r and z (in the units of $a_{HO}$) for atoms and molecules respectively, for different values of radial and axial

quantum numbers (*n and $n_z$*) for a fixed azimuthal quantum number *l*=2. It is found that with the increase in n, the number of nodes in the density profile along the r-axis increases for a fixed value of $n_z$. However with the increase in $n_z$ density profile (crest and trough structure) obtained for the particular value of n is replicated or repeated $n_z$ times along the z axis giving rise to total number of nodes along the r-axis, k= (n+1)($n_z$+1), including the node (minimum) at r=0, for a fixed value of l=2. It is to be noticed that due to this replication or repeatation along the z-axis, minima in the density profile are developed along the z-axis. The central vortex appears around r=0 line.

Corresponding 3D ring vortices for atoms and molecules have been plotted in Fig.2(c) and Fig.2(d) respectively. It is evident from these two figures that different combinations of radial (n) and axial ($n_z$) quantum numbers for a fixed value of azimuthal quantum number l=2 give rise to different structure of ring vortices and are centred at r=0 line. With the increase in the number of nodes and crest in the atomic density profile with the increase in n for a fixed value of $n_z$ (as shown in Fig.2(a)), more layers of crest and trough are added around the central vortex giving rise to different structure around the central vortices as shown in Fig.2(c). Similar feature has been obtained previously for atomic ring vortices for $n_z$ =0 by Li et al [7]. However with increase in $n_z$, the 3D ring vortex structure obtained for a fixed value of n is replicated or repeated ($n_z$+1)-fold in the vertical direction (centred at r=0 line). Fig.2(d) shows that the same structure of molecular ring vortices as that of atomic vortices appear for different combinations of n and $n_z$ and for l=2. Comparison of these two figures show that the molecular vortices are localized to smaller region in comparison to atomic vortices by an amount ( $1/\sqrt{2}$) due to greater mass which is twice of atom.

**IIIB. Evolution of Vortices:**

We studied the formation and evolution of atomic and molecular vortices by solving the time-dependent coupled equations (12) and (13) using Steepest-Descent method in Crank-Nicholson scheme with and without external decay term.

Fig.3 Illustrates the variations of the number of atoms and molecules in Fig.3a and Fig.3c respectively with time for x=1 and x =4 without any decay at two different combinations of radial and axial quantum numbers(i) $n=1, n_z=0$ and (ii) $n=3$, $n_z=2$ at fixed value of azimuthal quantum number $l=2$. These two figures show there exists coherences in the oscillation of atomic and molecular numbers in two different vortex states leading to out of phase oscillations of atomic and molecular numbers with time due to the presence of atom-molecular coupling in this coupled system. For larger intensity of lasers i.e. x=4 the oscillation frequency increases due to the increase in the Rabi-frequency between states coupled by the lasers $I_1$ and $I_2$, but the coherence is maintained in the atomic and molecular number oscillations. Furthermore, the amplitude of oscillation is large for lower vortex state (*n=1,n$_z$ =0*) for a fixed x (compare blue and red lines with olive and wine lines) both in the case of atomic and molecular vortices.

It is found that with the inclusion of external decay ( Fig.3b and Fig.3d), the nature of oscillations are the same and coherence is preserved, but the amplitude of oscillations die out much faster than that in absence of decay. The damping in amplitude of oscillations is larger in the higher vortex state (n=3, $n_z$=2) than that in the lower vortex state (n=1, $n_z$=0) both in atomic and molecular cases for both the values of x. Moreover for higher intensity (x=4) the damping in oscillation is large due to the increase in the induced decay with the increase in intensity for both the species in two vortex states. Therefore we show that the coherence in number oscillation is present in the evolution of rotating atom-molecular condensate system coupled by two-photon Raman photo-association in different vortex states, similar to that in the non-rotating coupled BEC in the ground state [21].

Fig.4 shows evolution of coupled atomic and molecular quantized 3D ring vortices in the vortex state for n=3, $l$=2, $n_z$=2 at different time without external decay (Fig.4a) and with external decay (Fig.4b) considering x=4 for both the cases. In figures (4a) and (4b) we have also shown how the number of atoms and molecules oscillates out of phase with time in this vortex state without and with external decay respectively. Comparing the set of figures for atomic vortices at different time with the corresponding molecular vortices at the same time it is found that the intensity of crests of the molecular vortices grows at the expense of that for atomic vortices and vice-versa. This feature is present for both the cases, without and with external decay. However faster decay of the intensity of the crests is found in coupled ring vortices in presence of external decay than that in absence of it. Therefore the signature of coherence is found to be implemented on the evolution of coupled atomic and molecular ring vortices.

**IIIC. Stability Analysis**:

The variations of conversion efficiency (η) of atoms into molecules and the life time (τ) of the system (time for the decay of total number of atoms to 1/e times its initial number) with x and the total number of atoms N in two vortex states in presence of external decays (both spontaneous and induced decay) are presented in Fig.5. Dependence of efficiency of conversion from atoms to molecules and the lifetime for two different vortex states as a function of x are shown in Fig.5a and Fig.5c respectively. For both the vortex states (*n*=1, *l*=2, $n_z$=0 and *n*=3, *l*=2, $n_z$=2) the variation of conversion efficiency and the lifetime with x show opposite nature, conversion efficiency sharply increases to the saturation whereas lifetime sharply decreases to small value with the increase in x. Conversion efficiency is saturated in the lower vortex states after x=8, while it starts to saturate from x=15 for higher vortex state. It is also found efficiency of conversion is lower for higher energy vortex state (*n*=3, *l*=2, $n_z$=2) which is separated by energy about 2.6 in unit of ℏω from the lower state. However the decay rate is the same for both the vortex states. Variation of conversion efficiency and the life time with N are shown in Fig.5(b) and Fig.5(d) respectively. Similar to the variation with x (Figs 5(a) and 5(c)), conversion efficiency increases and the lifetime decreases with increase in N. The rate of increase of conversion efficiency is higher for lower vortex state than that for higher vortex state (Fig.5(b)), whereas the rate of decrease of lifetime is higher for lower vortex state than that for higher vortex state with increase in N (Fig.5(d)).

Conversion efficiency curves (solid lines in Figs. 5(a) and 5(b)) is obtained by fitting the calculated points shown in the figures. In Fig.5(a) the fitted efficiency curve follows the relation with x as $\eta = A(1 - e^{-bx})$, where 'A' and 'b' are constants [Table I] and they depend on the initial number of atoms and the energy of the system. Furthermore, the variation of conversion efficiency with initial particle numbers follow the similar trend as it changes with x ($i.e. \eta = A(1 - e^{-bN})$) [Fig.5(b)]. The curve fitting of calculated data points shows the variation of the average life time of the particles with 'x' as $\tau = A/x^d$, [Fig.5(c)] where d is slightly greater than unity (Table-I). But the lifetime of the system falls linearly with N (τ=A+bN) [Fig.5(d)] unlike its variation with x. The tabulated fitting parameters in Table-I shows that for η vs x and η vs N variations the constant A is lower for higher vortex state while the constant b is twice for the lower state than that of higher state. For τ vs x variation of both the constants A and b are same for both the lower and higher states. For τ vs. N variation constant A is almost the same for both the lower and upper states and the value of b is larger for lower vortex state.

In the study of formation and evolution of coupled atomic and molecular vortices in the rotating atom-molecular BEC system demands investigation on the stability of such system. It is known that the stability of the atomic and molecular vortices are inversely proportional to the value of imaginary part of the energy of atomic and molecular vortex states $Im(E_{a,m})$. To demonstrate the stability of atomic and molecular vortices in rotating coupled system we have plotted $Im(E_{a,m})$ for two different vortex states (Fig.6). Fig.6 ((a) - (d)) provide the variation of $Im(E_{a,m})$ for both the atomic and molecular vortex states with respect to x, N, $\lambda_{am}/\lambda_a$ and $\Omega/\omega_r$ respectively. In the stability analysis effect of external decay has been considered. Variation of $Im(E_{a,m})$ with x (fig.6a) shows that atomic vortices in both the energy states are almost stable. But the stability of molecular vortices decreases sharply in the lower energy state than that in the higher energy state. It is shown in Figs.5(a) and 5(c) that the conversion efficiency and the lifetime behave oppositely with increase in x and hence it will contribute to the energy of the system oppositely. With the increase in x the rate of increase of conversion efficiency i.e. the rate of formation of molecules is higher for the lower vortex state whereas the rate of decay of the system is the same for both the states. By repeating the calculation by using lower value of conversion efficiency (1/4 of its original value) we found that the stability of the lower state increases (not shown here). Thus higher value of conversion efficiency of lower state may give rise to the instability for the lower molecular vortex state higher than that for the upper molecular vortex state when the decay rate of both the states are the same.

Fig.6b shows that atomic vortices are stable with N whereas stability of molecular vortices decreases with the increase in total number of particles and the stability is much less than that of the atomic vortices. Dependence of stability on the strength of atom-molecular interaction in the units of atom-atom interaction shows (Fig.6(c)) atomic vortices are stable for both the vortex states although that in the higher energy state increases slowly with increase in atom-molecular interaction. However the stability of molecular vortices is much less than that of atomic vortices and it becomes more unstable with increase in atom-molecular interaction. The stability of molecular vortices in the lower state is greater than that of the higher state by a factor of two in the range $\lambda_{am} \leq \lambda_a$. Similarly increase in rotation frequency in units of trap frequency (Fig.6(d)) shows stable atomic vortices and the stability is much more than that of the molecular vortices. The stability of molecular vortices in the lower state is twice greater than that in the higher state in the range $\Omega \leq \omega_r$. These figures reveal that although atomic vortices are stable in both the energy states and more or less independent of system parameters (x, N, $\lambda_{am}/\lambda_a$ and $\Omega/\omega_r$), stability of molecular vortices is less than that of atomic vortices and it decreases with the increase in these system parameters except for rotation frequency. With increase in $\Omega/\omega_r$ molecular stability slightly increases. Moreover the stability in the higher energy vortex state is lower than that in the lower energy vortex state except in the variation with x. Therefore the stability of atomic and molecular ring vortices in coupled rotating atom-molecular BEC system can be controlled by varying different system parameters such as intensity of PA lasers, initial number of atoms, interaction strengths and the rotation frequency of the system.

## IV. Conclusion:

A detailed investigations on the rotating coupled atom-molecular Bose-Einstein Condensates of $^{87}$Rb trapped in a 3D-anisotropic cylindrical trap in both time-independent and time-dependent Gross-Pitaevskii approaches reveal that coupled atomic and molecular quantized ring vortices can be achieved in coupled BEC system. For different combinations of radial (*n*) and axial (*n_z*) quantum numbers at a fixed azimuthal quantum number (*l*) different number of nodes are obtained in density profile (as a function r and z). The presence of these nodes and crests in the density profile give rise to different structure around the central vortex at r=0. Although the structure of vortices formed for a particular combination of n, $n_z$ and *l* are the same for atomic and molecular vortices, the nature of the vortices (i.e. the spread and the peak density of crests) are different. Molecular vortices are more localized than atomic vortices. Evolution of population of atoms and molecules with time in two vortex states shows out of phase oscillations in both the cases, with and without external decays. However the amplitude of oscillation decays fast in presence of external decay. The out of phase oscillation of atomic and molecular numbers in this rotating coupled BEC system shows signature of coherences as par with the coherences present in the evolution of atomic and molecular numbers in the ground state in absence of any vortices. This coherence is implemented also on the evolution of coupled atomic and molecular ring vortices as their intensity increases or decreases oppositely in the rotating coupled BEC system both in presence and absence of external decay. To investigate the feasibility and stability of the atomic and molecular vortices in the rotating coupled BEC system we have studied the lifetime of the system and atom to molecular conversion efficiency as a function of laser intensity and the initial number of atoms, considering the external decay. It is found from the life time variation that although the system decays sharply with increase in laser intensity, life time of the system decreases slowly with increase in initial number of atoms. Variation of conversion efficiency shows it is saturated with increase in laser intensity and with increase in initial number of atoms it increases steadily. Therefore it can be stated that it is feasible to design coupled atomic and molecular ring vortices in the rotating BEC system depending on the choice of laser intensity and the initial number of atoms. Linear stability analysis reveals that the stability of the coupled atomic and molecular ring vortices can be controlled by changing the system parameters e.g. laser intensity, initial number of atoms, interaction strength and rotation frequency of the system. It is found that the atomic vortices are stable with the variation of these parameters, but the molecular vortices are less stable than the atomic vortices. Therefore stability of atomic and molecular vortices can be controlled with the optimum choice of the system parameters.

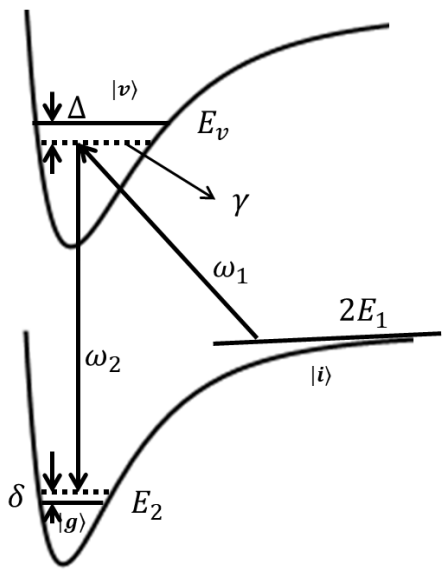

Fig.1:Schematic representation of two-photon Raman photoassociation

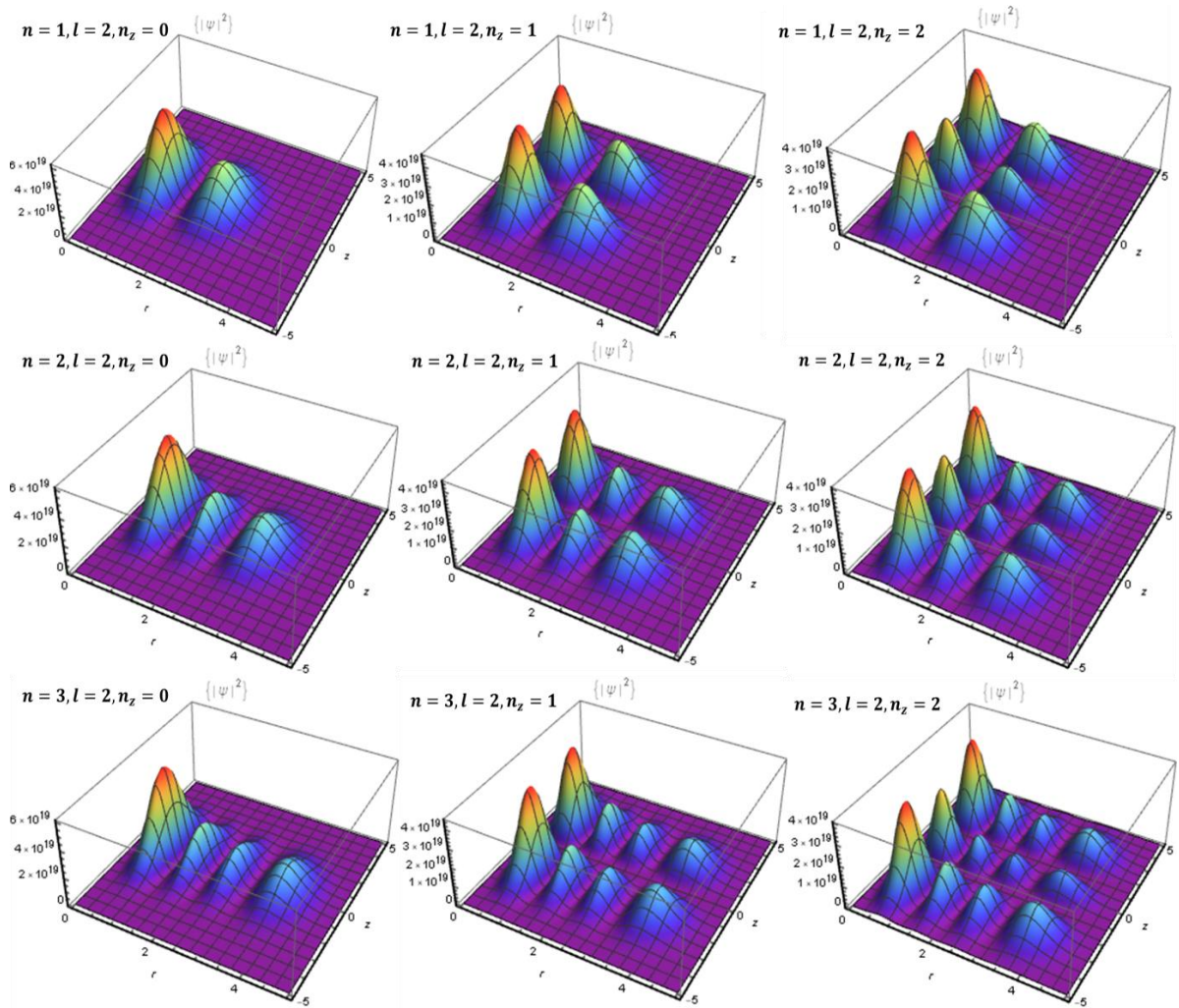

Fig.2(a): Atomic density $|\Phi_a(r,z)|^2$ of the stationary state vortex solution for three dimensional coupled BEC system as a function of radial distance r (unit of $a_{HO}$) and axial distance z (unit of $a_{HO}$) for different combination of radial (n) and axial ($n_z$) quantum numbers keeping azimuthal quantum numbers ($l$) fixed at 2.

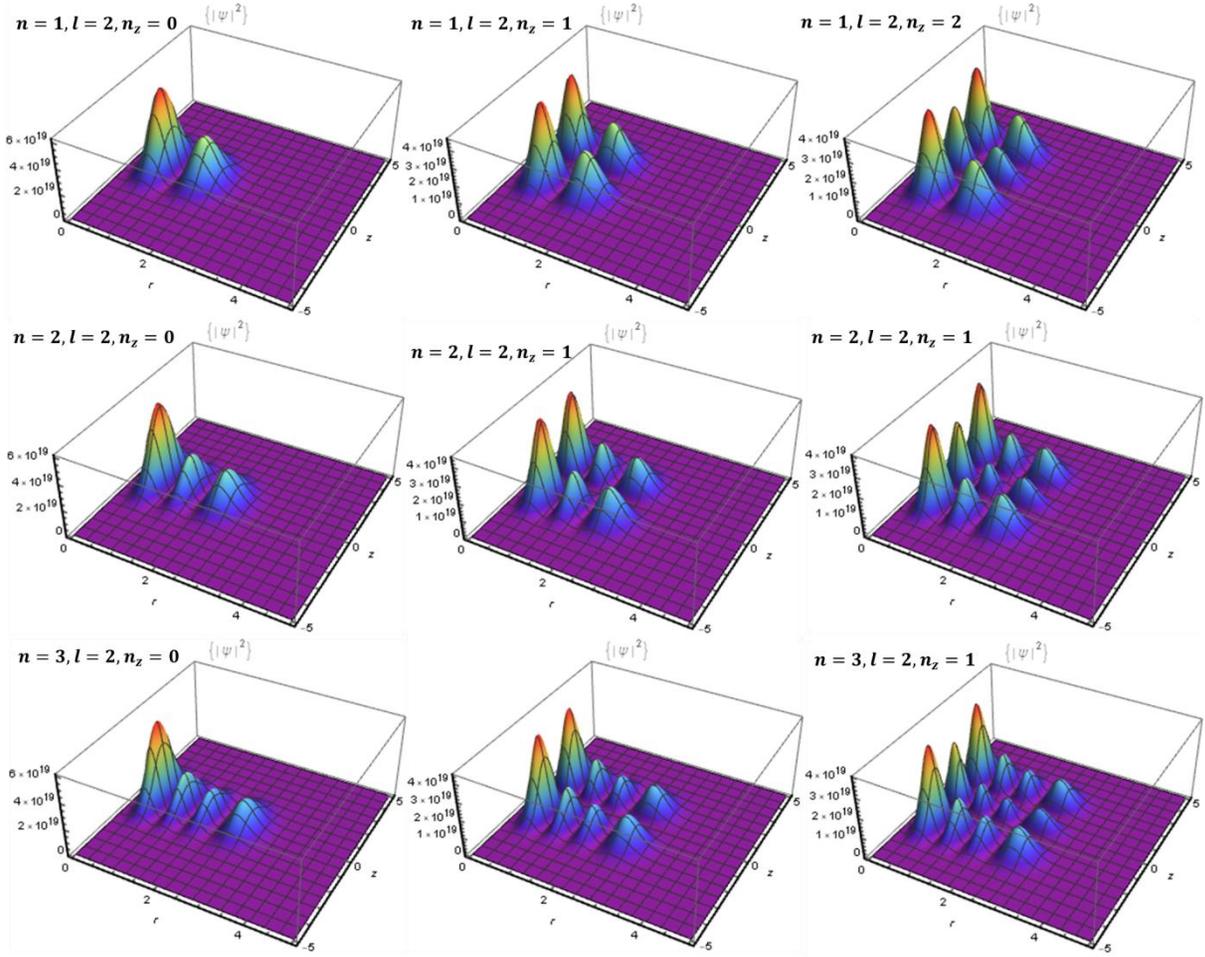

Fig.2(b): Molecular density $|\Phi_m(r,z)|^2$ of the stationary state vortex solution for three dimensional coupled BEC system as a function of radial distance r (unit of $a_{HO}$) and axial distance z (unit of $a_{HO}$) for different combination of radial (n) and axial ($n_z$) quantum numbers keeping azimuthal quantum numbers ($l$) fixed at 2.

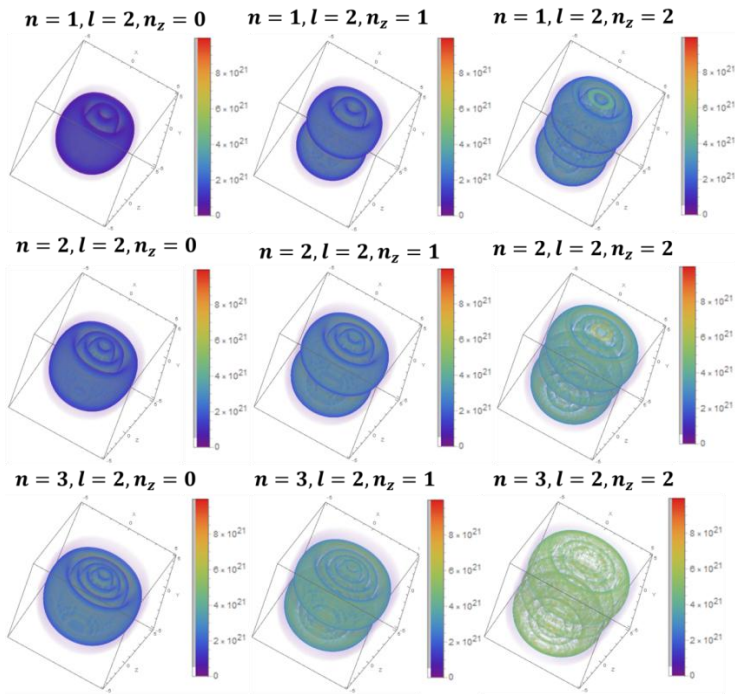

Fig.2(c): Atomic Vortices for different combinations of radial (n) and axial ($n_z$) quantum numbers for a fixed azimuthal quantum number l=2

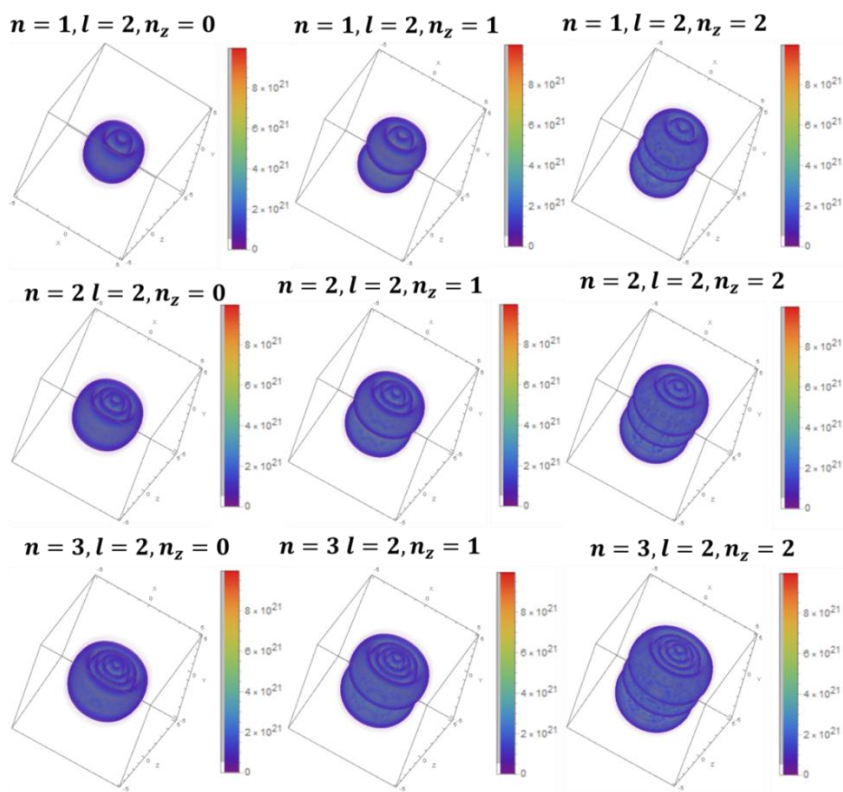

Fig.2(d): Molecular Vortices for different combinations of radial (n) and axial ($n_z$) quantum numbers for a fixed azimuthal quantum number l=2

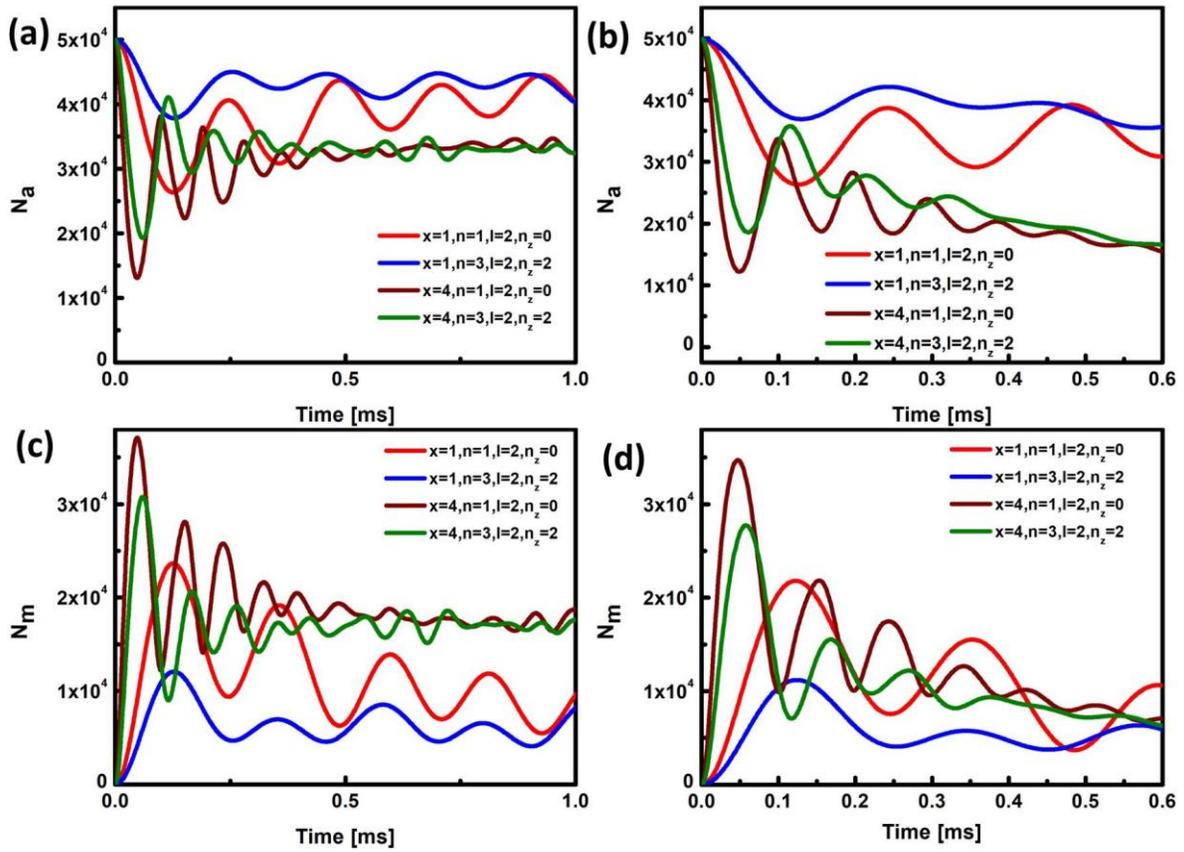

Fig.3: Variation of atomic (a) and molecular (c) number with time for x=1 and x=4 with decay=0 for two different combination of radial quantum number (n=1,3) and axial quantum number ($n_z$=0,2) with azimuthal quantum number fixed at *l*=2 and the variation of atomic (b) and molecular (d) numbers with time for the same combination of n, *l* and $n_z$ with external decay.

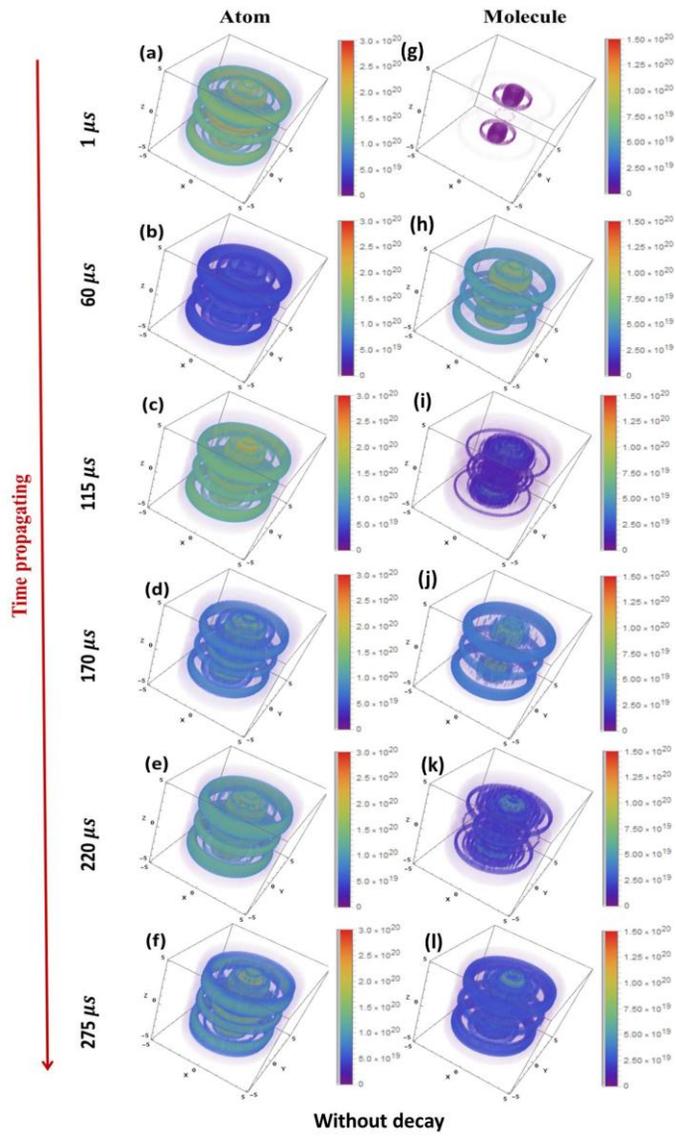

Without decay

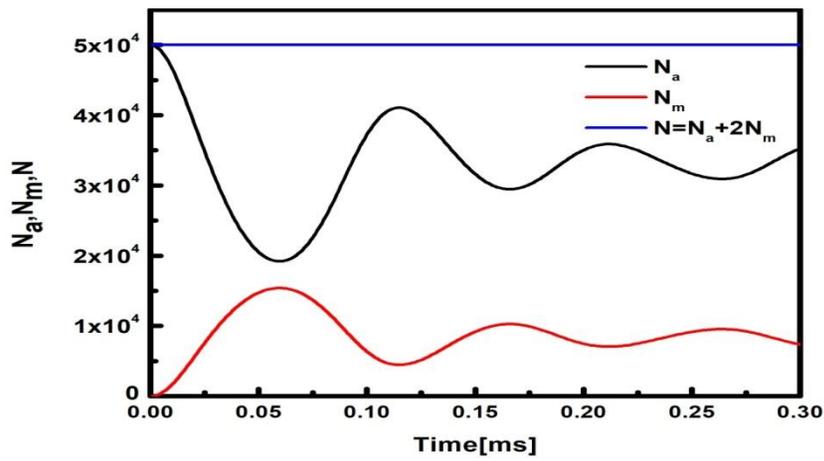

Figure 4(a)

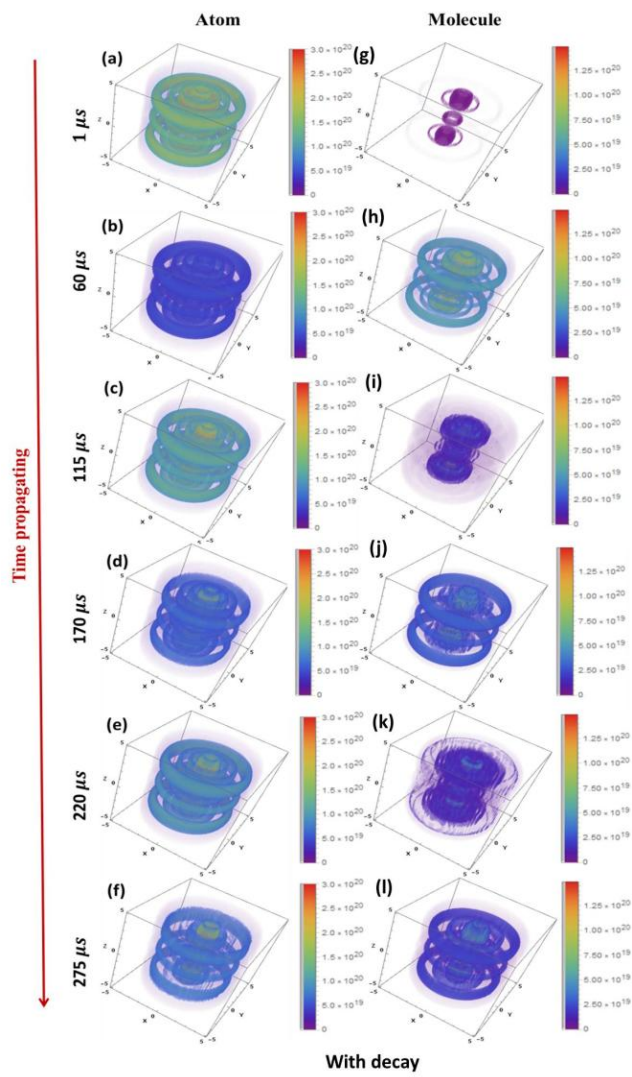

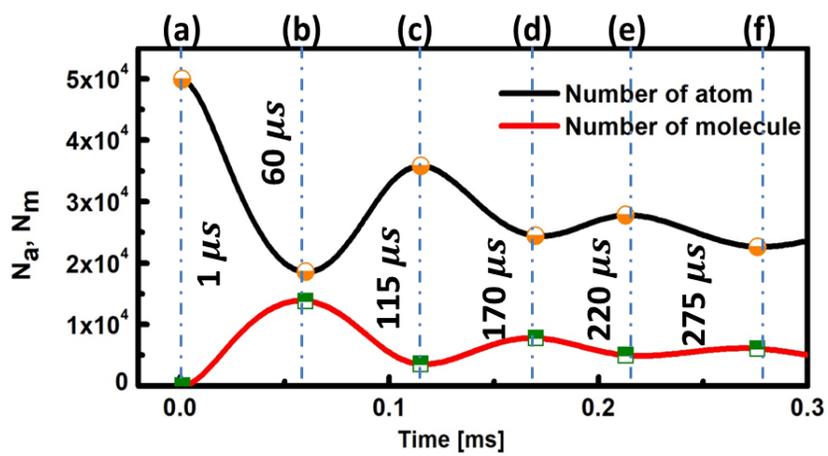

Figure: 4(b)

Fig.4: Evolution of coupled atomic and molecular 3D ring vortices in the vortex state for n=3, l=2, $n_z$=2 at different times (a) without and (b) with external decay for atom (on the left) and for molecule (on the right) for x=4 and N=50000. Evolution of number of atoms and molecules with time for n=3, l=2, $n_z$=2 vortex state has also been shown below that of the corresponding ring vortices.

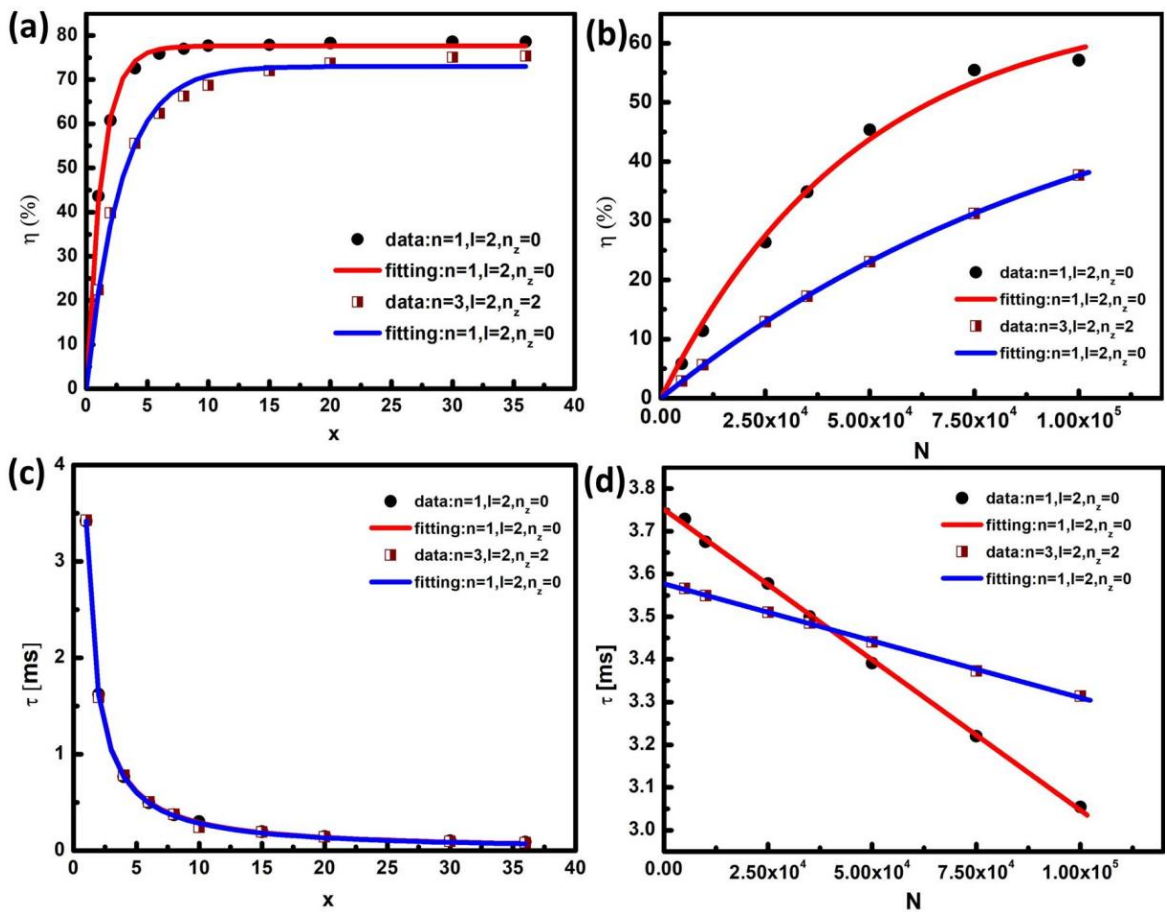

Fig.5: Variation of (a) Conversion efficiency of atoms into molecules and (c) Average Life time with *x* (light intensity factor) for two different vortex states

$n=1$, $l=2$, $n_z=0$ (red line) and $n=3$, $l=2$, $n_z=2$ (blue line). Variation of (b) Conversion Efficiency and (d) Life Time, for vortex states with the quantum numbers $n=1$, $l=2$, $n_z=0$ (red) and $n=3$, $l=2$, $n_z=2$ (blue) for $x=1$, with total number of particle (N). Here $\lambda_a = \lambda_m = \lambda_{am}$. Results of calculations are given as points and the lines are from curve fittings.

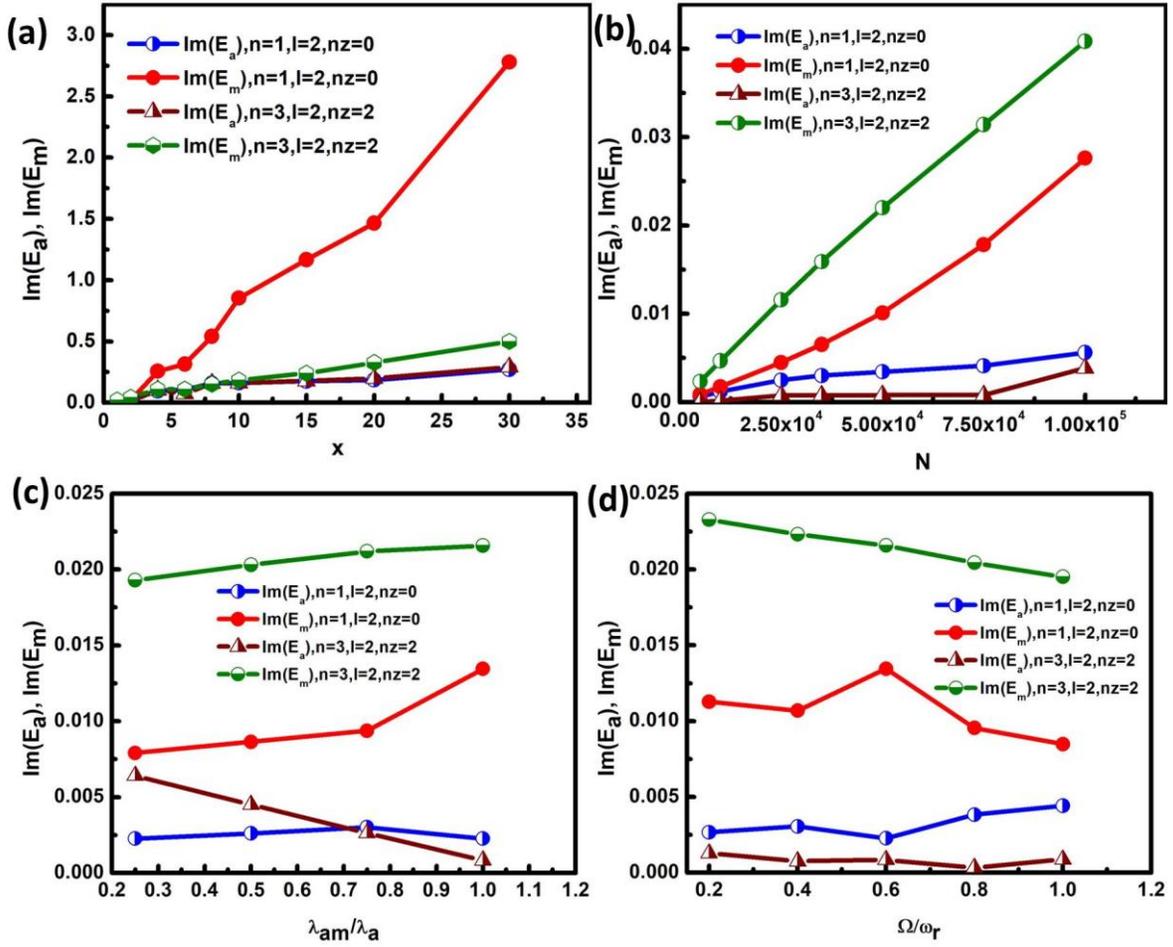

Fig.6: Stability curve for atom (blue and wine line) and molecule (red and olive line) with the variation of (a) x, (b) N, (c) relative interaction strengths $\lambda_{am}/\lambda_a$ and (d) external rotating frequency to trap frequency $\Omega/\omega_r$ for vortex states with $n=1$, $l=2$, $n_z=0$ (blue and red lines) and $n=3$, $l=2$, $n_z=2$ (wine and olive line). Here $x=1$ for N, $\lambda_{am}/\lambda_a$ and $\Omega/\omega_r$ variation curves.

Table I: Fitting parameters for the conversion efficiency (η) and average lifetime (τ) of the system with x and N

| X variable | Y variable | Energy level | Parameter(1) | Parameter(2) |
|---|---|---|---|---|
| x | η | $n=1, l=2, n_z=0$ | A=77.62 | b=0.784 |
| x | η | $n=3, l=2, n_z=2$ | A=73.01 | b=0.355 |
| N | η | $n=1, l=2, n_z=0$ | A=67.33 | $b=2.10 \times 10^{-5}$ |
| N | η | $n=3, l=2, n_z=2$ | A=62.33 | $b=9.26 \times 10^{-6}$ |
| x | τ | $n=1, l=2, n_z=0$ | A=3.42 | d=1.069 |
| x | τ | $n=3, l=2, n_z=2$ | A=3.42 | d=1.082 |
| N | τ | $n=1, l=2, n_z=0$ | A=3.75 | $b=-7.04 \times 10^{-6}$ |
| N | τ | $n=3, l=2, n_z=2$ | A=3.58 | $b=-2.67 \times 10^{-6}$ |